\documentclass[preprint,notoc]{JHEP3}
\usepackage{epsfig}

\def\to{\rightarrow}

\def\bi{\begin{itemize}}
 \def\ei{\end{itemize}}

\def\c1p{C1^\prime}
\def\ta{\tilde a}
\def\tG{\tilde G}

\def\ta{\tilde a}

\def\tst{\tilde t}

\def\tw{\tilde\chi}

\def\tz{\tilde\chi^0}
\def\alt{\stackrel{<}{\sim}}
\def\agt{\stackrel{>}{\sim}}
\def\be{\begin{equation}}  
\def\ee{\end{equation}}  
\def\bea{\begin{eqnarray}}  
\def\eea{\end{eqnarray}}

\newcommand\sjp[3]{{\it Sov.\ J.\ Nucl.\ }{\bf #1} (#2) #3}

\title{
Neutralino versus axion/axino cold dark matter\\
in the 19 parameter SUGRA model
}
\author{Howard Baer$^{a}$, Andrew D. Box$^{a}$ and Heaya Summy$^{a}$\\
$^a$Dept.\ of Physics and Astronomy, University of Oklahoma, Norman, OK 73019, USA\\
E-mail: \email{baer@nhn.ou.edu}, \email{box@nhn.ou.edu},\email{heaya@nhn.ou.edu}}

\preprint{\vbox{}}

\abstract{
We calculate the relic abundance of thermally produced neutralino cold dark
matter in the general 19 parameter supergravity (SUGRA-19) model. 
A scan over GUT scale parameters reveals that
models with a bino-like neutralino typically give rise to a dark matter
density $\Omega_{\tz_1}h^2\sim 1-1000$, {\it i.e.} between 1 and 4 orders of
magnitude higher than the measured value. Models with higgsino or
wino cold dark matter can yield the correct relic density, but mainly
for neutralino masses around 700-1300 GeV. 
Models with mixed bino-wino or bino-higgsino CDM,
or models with dominant co-annihilation or $A$-resonance annihilation
can yield the correct abundance, but such cases are extremely hard to 
generate using a general scan over GUT scale parameters; this is indicative of
high fine-tuning of the relic abundance in these cases. 
Requiring that $m_{\tz_1}\alt 500$ GeV (as a rough naturalness requirement) 
gives rise to a minimal probably dip in parameter space at the measured CDM abundance.
For comparison, we also scan over mSUGRA space with four free parameters.
Finally, we investigate the Peccei-Quinn augmented MSSM 
with mixed axion/axino cold dark matter.
In this case, the relic abundance agrees more naturally with the
measured value.
In light of our cumulative results, 
we conclude that future axion searches should probe
much more broadly in axion mass, and deeper into the axion coupling. 
}

\begin{document}

\section{Introduction}
\label{sec:intro}

The cosmic abundance of cold dark matter (CDM) has been recently extracted
at high precision by the WMAP collaboration\cite{wmap7}. An analysis of 
WMAP seven year data implies
\be
\Omega_{CDM}h^2=0.1123\pm 0.0035\ \ \ (68\%\ CL) ,
\ee
where $\Omega=\rho/\rho_c$ is the dark matter density relative to the 
closure density, and $h$ is the scaled Hubble constant.
Since no particle contained within the Standard model of particle physics
has exactly the right properties to constitute CDM, new matter states
from physics beyond the SM are required for explanation\cite{dmreview}. 

The most popular explanation of CDM comes from models including weak scale
supersymmetry (SUSY)\cite{wss}. SUSY models are invoked to cure the hierarchy problem
contained within the SM. SUSY models also receive some indirect support 
from experiment\cite{gauge} in that the gauge couplings, measured at the scale $Q=M_Z$,
meet nearly at a point when run to $M_{GUT}$ via the renormalization group
equations (RGEs) of the Minimal Supersymmetric Standard Model, or MSSM\cite{drw}.
Within $R$-parity conserving SUSY theories, 
there are several dark matter candidates, including
the lightest neutralino $\tz_1$, the gravitino $\tG$ and the
axion/axino admixture $a\ta$ from the Peccei-Quinn\cite{pq,ww} 
augmented MSSM (or PQMSSM).\cite{steffen_review}

Most analyses focus upon the lightest neutralino $\tz_1$ as a CDM
candidate, since it can be classified as a weakly interacting massive
particle, or WIMP. WIMP particles such as $\tz_1$ are especially compelling
due to the so-called ``WIMP miracle''. It is assumed the DM particle
is present in thermal equilibrium at some early time in the universe's 
history, and then the relic abundance can be found by solving the
Boltzmann equation for its number density. Particles with weak scale
mass, which interact with weak scale strength, should give a relic abundance
roughly in the vicinity of what is measured.

However, detailed calculations of the neutralino relic abundance
in models such as mSUGRA\cite{msugra} (or CMSSM) find that throughout parameter space, 
the neutralino abundance is usually much higher than the measured value, 
and typically by 1-2 orders of magnitude\cite{inoDM}.
Only in very narrow regions of parameter space is the measured
CDM abundance found: the stau co-annihilation region\cite{stau}, the hyperbolic branch/
focus point region with mixed higgsino DM\cite{hb_fp}, the $A$-resonance region
at large $\tan\beta$\cite{Afunnel}, and perhaps at the light Higgs resonance\cite{higgs}. 

That the mSUGRA parameter regions where the predicted abundance agrees with
measurement are quite narrow is indicative of a needed fine-tuning of
fundamental model parameters to achieve the required value.
The fine-tuning is a measure of the slope of the surface of 
$\Omega_{\tz_1}h^2$ as a function of model parameters $a_i$.

Quantitatively, the fine-tuning with respect to variation in the
parameter $a_i$ is defined as
\begin{equation}
\Delta_{a_i} \equiv \frac{\partial\log\Omega_{DM}h^2}{\partial\log a_i} .
\end{equation}
The overall fine-tuning can be obtained by combining the
$\Delta_{a_i}$ in quadrature. The degree of fine-tuning of the
relic abundance in mSUGRA has been calculated in Ref's \cite{eo}
and \cite{bbox}. Both groups find that the regions of agreement
between theory and experiment are highly fine-tuned. Ref. \cite{bbox}
claims also that the very large $\tan\beta$ ($\agt 50$) region 
is not very fine-tuned, 
since the pseudoscalar Higgs width $\Gamma_A$ becomes very large
(owing to the large $b$-quark Yukawa coupling), and efficient
neutralino annihilation can occur all over the $m_0\ vs.\ m_{1/2}$
plane. 

In Ref. \cite{bbs}, we proposed mixed axion/axino cold dark matter
in the Peccei-Quinn augmented mSUGRA model.
By requiring a re-heat temperature in the
$10^6-10^8$ GeV range, a MeV-scale axino mass and a large value of 
$f_a\agt 10^{11}$ GeV is favored.
As opposed to the high fine-tuning which is present in
mSUGRA for neutralino CDM, if instead mixed axion/axino CDM is invoked, 
with a $\sim$ MeV scale axino\cite{axmass}, then the fine-tuning needed to achieve
a theory-experiment match is relatively low. The lowest fine-tuning is observed
for the case where there is a nearly equal mix of axion and thermally produced
axino DM\cite{bbox}.

A major criticism of the result found in Ref. \cite{bbox} is that
the fine-tuning analysis of neutralino CDM is restricted to the mSUGRA model. 
It might be possible that if one opens up the parameter space to include models with non-universality, 
then many more possibilities may arise to gain a theory-experiment match in the
dark matter relic density.

In this paper, we calculate the relic abundance of neutralino CDM 
in a 19 parameter version of the MSSM: the SUGRA-19 model. 
Another recent calculation of the neutralino CDM abundance in a 19 parameter
MSSM has been performed by Berger {\it et al.}\cite{bghr}.\footnote{
Similar analyses using weak scale SUSY parameters and requiring a thermal
abundance of neutralino CDM in accord with WMAP are given in 
Ref's \cite{py,aaqfh,bbps}.} 
In that analysis, 
the so-called pMSSM model is examined, which is a version of the MSSM with
minimal $CP$ and flavor violating terms. 
The 19 soft SUSY breaking and other parameters are
input at the weak scale, and it is required that $\Omega_{\tz_1}h^2\le 0.11$.
In the analysis presented here, we will also adopt a 19 parameter MSSM, but will insist
on adopting soft SUSY breaking parameters {\it at the GUT scale}. 
The reason is that the unification of gauge couplings provides 
circumstantial evidence that the MSSM, 
or MSSM plus gauge singlets (or additional $SU(5)$ multiplets) is the correct 
effective field theory between $M_{weak}$ and $M_{GUT}$. The pattern of soft term running
as provided by the MSSM RGEs gives us some guidance as to how likely  are various sets of
weak scale parameters. For instance, by sampling over weak scale parameters including
the superpotential $\mu$ term, one easily generates low $\mu$ solutions which lead
to higgsino-like neutralinos. However, sampling over GUT scale parameters, it is not 
easy to generate low $\mu$ solutions with higgsino-like dark matter. The reason is that
(at tree level) $\mu^2(m_{weak})\simeq -m_{H_u}^2(M_{weak})$, and it is rare that $m_{H_u}^2$ 
runs to just barely negative values, which would lead to a small $\mu$ parameter.
In addition, in distinction with Ref. \cite{bghr}, we will not restrict ourselves to 
requiring $\Omega_{\tz_1}h^2<0.11$, but will instead allow all values of thermal neutralino abundance. 

In Sec. \ref{sec:method}, we provide an overview of our scanning methodology and
parameter space values, and our calculation of the neutralino relic abundance.
In Sec. \ref{sec:wimp}, we present results of the neutralino relic density from a 
scan over  the SUGRA-19 model. We find that solutions with bino-like
neutralinos give a large overabundance of CDM. Models with higgsino-like or
wino-like neutralinos tend to give too little CDM, unless $m_{\tz_1}\sim 1$ TeV, with
observable sparticles likely beyond LHC reach. The value $\Omega_{\tz_1}h^2\sim 0.11$
lies at the minimum probability between these two extremes. Since our scan depends on
the bias as to how we sample the GUT scale parameters, we present results for 
both a linear as well as a log scan; the latter scan accentuates small soft 
SUSY breaking parameter values.
In Sec. \ref{sec:sugra}, we present for comparison a similar analysis done for the
mSUGRA model. 
In Sec. \ref{sec:axino}, we work instead within the Peccei-Quinn augmented MSSM, or PQMSSM,
which leads to mixed axion/axino CDM if the axino is assumed to be the lightest SUSY particle (LSP).
The PQMSSM requires four additional parameters: the PQ breaking scale $f_a$ and initial mis-alignment
angle $\theta_i$, the axino mass $m_{\ta}$, 
and the re-heat temperature of the universe $T_R$.
In this case, the PQMSSM with 23 free parameters can more easily yield
the measured relic abundance. We evaluate the favored range of the additional PQMSSM parameters.
Our conclusions are presented in Sec. \ref{sec:conclude}.

\section{Relic density in the SUGRA-19 model}
\label{sec:method}

For our calculations, we adopt the Isajet 7.80\cite{isajet,bfkp} SUSY spectrum generator Isasugra.
Isasugra begins the calculation of the sparticle mass spectrum with
input $\overline{DR}$ gauge couplings and $f_b$, $f_\tau$ Yukawa couplings at the 
scale $Q=M_Z$ ($f_t$ running begins at $Q=m_t$) and evolves the 6 couplings up in energy 
to scale $Q=M_{GUT}$ (defined as the value $Q$ where $g_1=g_2$) using two-loop RGEs. 
At $Q=M_{GUT}$, the SSB boundary conditions are input, and  the set
of 26 coupled two-loop MSSM RGEs\cite{mv} are evolved back down in scale to
$Q=M_Z$. 
Full two-loop MSSM RGEs are used for soft term evolution, while the gauge and Yukawa coupling
evolution includes threshold effects in the one-loop beta-functions, so the gauge and
Yukawa couplings transition smooothly from the MSSM to SM effective theories as 
different mass thresholds are passed.
In Isajet 7.80, the values of SSB terms which mix are frozen out at the
scale $Q\equiv M_{SUSY}=\sqrt{m_{\tst_L} m_{\tst_R}}$, while non-mixing SSB terms are frozen out
at their own mass scale\cite{bfkp}. 
The scalar potential is minimized using the RG-improved one-loop MSSM effective
potential evaluated at an optimized scale $Q=M_{SUSY}$ which accounts for
leading two-loop effects\cite{haber}.
Once the tree-level sparticle mass spectrum is computed, full one-loop
radiative corrections are caculated for all sparticle and higgs boson masses,
including complete one-loop weak scale threshold corrections for the
top, bottom and tau masses at scale $Q=M_{SUSY}$. These fermion self-energy 
terms are critical to evaluating whether or not Yukawa couplings do
indeed unify.
Since the GUT scale Yukawa couplings are modified by the threshold corrections, the
Isajet RGE solution must be imposed iteratively with successive up-down 
running until a convergent sparticle mass solution is found.
For most of parameter space, there is excellent agreement between Isajet and
the SoftSUSY, SuSpect and Spheno codes, although at the edges of parameter
space agreement between the four codes typically diminishes\cite{kraml}.
We adopt the Isasugra non-universal SUGRA parameter space.

We will implement at first a linear scan over the following parameters.
\begin{itemize}
\item Gaugino masses: $M_1,\ M_2,\ M_3: 0-3.5$ TeV
\item First/second generation scalar masses: $m_{Q_1}$, $m_{U_1}$, $m_{D_1}$,
$m_{L_1}$, $m_{E_1}$: $0-3.5$ TeV,
\item Third generation scalar masses: $m_{Q_3}$, $m_{U_3}$, $m_{D_3}$,
$m_{L_3}$, $m_{E_3}$: $0-3.5$ TeV,
\item Higgs soft masses: $m_{H_u},\ m_{H_d}:\ 0-3.5$ TeV,
\item trilinear soft terms: $A_t,\ A_b,\ A_\tau$:$-3.5\ {\rm TeV}\ \to 3.5$ TeV,
\item ratio of weak scale Higgs vevs $\tan\beta :\ 2-60$.
\end{itemize}
We adopt a common mass for first and second generation scalars so as to avoid
SUSY FCNC processes.

To gain an acceptable sparticle mass solution, we will require:
\begin{enumerate}
\item the lightest SUSY particle (LSP) is the neutralino $\tz_1$,
\item the lightest chargino, if non-wino-like, obeys the LEP2 limit $m_{\tw_1}>103.5$ GeV,
\item the lightest chargino, if wino-like, obeys the LEP2 limit 
$m_{\tw_1}>91.9$ GeV,
\item the light Higgs mass obeys the LEP2 limit $m_h>111$ GeV (where we 
allow a roughly 3 GeV uncertainty in the theory calculation as applied to
the actual limit where $m_h> 114.4$ GeV.
\end{enumerate}

For each acceptable solution, we calculate the neutralino relic density
$\Omega_{\tz_1}h^2$ using the IsaReD\cite{isared} program. IsaReD 
calculates all relevant neutralino annihilation and 
co-annihilation reactions, as obtained using CalcHEP, 
and then calculates the relativistic thermally-averaged 
(co)-annihilation cross sections times relative velocity.
Once the freeze-out temperature is determined, then the relic
density at the present time is found by integrating the Boltzmann equation 
as formulated for a FRW universe.

Since we assume the neutralino to be in thermal equilibrium, 
our relic density results do not explicitly depend on the value of the re-heat temperature
of the Universe $T_R$ after inflation. However, we must assume $T_R>T_{fo}\sim m_{\tz_1}/20$
so that $T_R$ is above the neutralino freeze-out temperature. Further, if $T_R\agt 10^{10}$ GeV, 
then thermal production of gravitinos in the early Universe, followed by decays to the LSP, 
will overproduce neutralino dark matter. Hence, for our neutralino CDM 
relic density calculations, we must assume here that
\bi
\item $m_{\tz_1}/20\alt T_R\alt 10^{10}$ GeV.
\ei

\section{Results for neutralino cold dark matter}
\label{sec:wimp}

\subsection{Linear scan over SUGRA-19 parameters}

Our first results from a linear scan over the above SUGRA-19 parameter
space is shown in Fig. \ref{fig:scan}, in the $\Omega_{\tz_1}h^2\ vs.\ 
m_{\tz_1}$ plane.\footnote{A qualitatively similar plot was generated
by Gelmini {\it et al.}\cite{ggsy} (Fig. 7) using an MSSM model with 9
input weak scale parameters.} 
The various solutions are color coded according to
the gaugino/higgsino content of the neutralino.
In the notation of Ref. \cite{wss}, 
if the bino-component $|v_4^{(1)}|>0.9$, then the neutralino is labeled
as bino-like (blue diamonds); if the wino-component $|v_3^{(1)}|>0.9$, then
it is labeled wino-like (purple $\times$); if the higgsino components
$\sqrt{v_1^{(1)2}+v_2^{(1)2}}>0.9$, then it is labeled as higgsino-like
(red squares). If the neutralino falls into none of these categories,
then it is labeled as ``mixed'' DM: (orange circles). 

We see from Fig. \ref{fig:scan} that the bino-like neutralinos
tend to populate the region with $\Omega_{\tz_1}h^2\gg 0.1$, {\it i.e.} 
usually about 2-3 orders of magnitude too high. For low values of
$m_{\tz_1}$, the abundance tends to be more like 3-5 orders of magnitude too high. 
A few bino-like points do tend to make it into the $\Omega_{\tz_1}h^2\sim 0.1$ region; 
these solutions tend to come form various co-annihilation 
or resonance annihilation processes. 
To obtain the required relic abundance via co-annihilation,
the LSP-NLSP mass gap must be tuned to just the right value. 
To obtain the required relic abundance via resonance annihilation,
the LSP mass must be adjusted to be close to half the mass of the 
resonance.
These co-annihilation and resonance annihilation points
are quite hard, but not impossible, to generate using our random scan
over GUT scale parameters.

The higgsino-like and wino-like CDM bands also show up as distinct
lines, typically with $\Omega_{\tz_1}h^2$ too low by 1-2 orders of magnitude
unless $m_{\tz_1}\agt 800-1200$ GeV. The wino-like band is relatively 
well-populated, as this just requires $M_2$ to be the lightest of the 
gaugino masses at the weak scale. The higgsino-like band is relatively
less populated, showing that higgsino-like CDM is rather fine-tuned
if one starts with GUT scale parameters, as mentioned in Sec. \ref{sec:intro}.
The points with the lowest population are those with mixed
bino-higgsino-wino CDM. These ``well-tempered neutralino\cite{wtn}'' 
points most naturally tend to populate
the $\Omega_{\tz_1}h^2\sim 0.1$ line, but they do require a fine-tuning
to avoid a bino, wino or higgsino dominance. Especially at low $m_{\tz_1}$,
relatively few solutions are found with $\Omega_{\tz_1}h^2\sim 0.1$. 
\FIGURE[t]{
\includegraphics[angle=0,width=12cm]{omgh2.eps}
\caption{Thermal abundance of neutralino cold dark matter from a 
linear scan over
the SUGRA-19 parameter space. We plot versus the neutralino mass. 
Models with mainly bino, wino, higgsino or a mixture are indicated by the 
various color and symbol choices. There are 5252 points in the figure.
}\label{fig:scan}}

To apprehend more clearly the dark matter probability distribution after our
linear scan of SUGRA-19 parameter space, we project the model points listed in
Fig. \ref{fig:scan} as a histogram onto the $\Omega_{\tz_1}h^2$ axis in
Fig. \ref{fig:bar1}{\it a}). Here we see the most probable value of $\Omega_{bino}h^2$ is
$\sim 10-100$ for bino-like dark matter (blue histogram), while the most probable
value for wino-like dark matter is $\Omega_{wino}h^2\sim 0.005-0.05$. The dip between these
two cases is partially filled in by cases of bino, higgsino or wino, or a mixture,
with the minimum probability lying around $\Omega_{\tz_1}h^2\sim 0.2-0.4$,
{\it i.e.} just above the measured value. 
\FIGURE[t]{
\includegraphics[angle=-90,width=12cm]{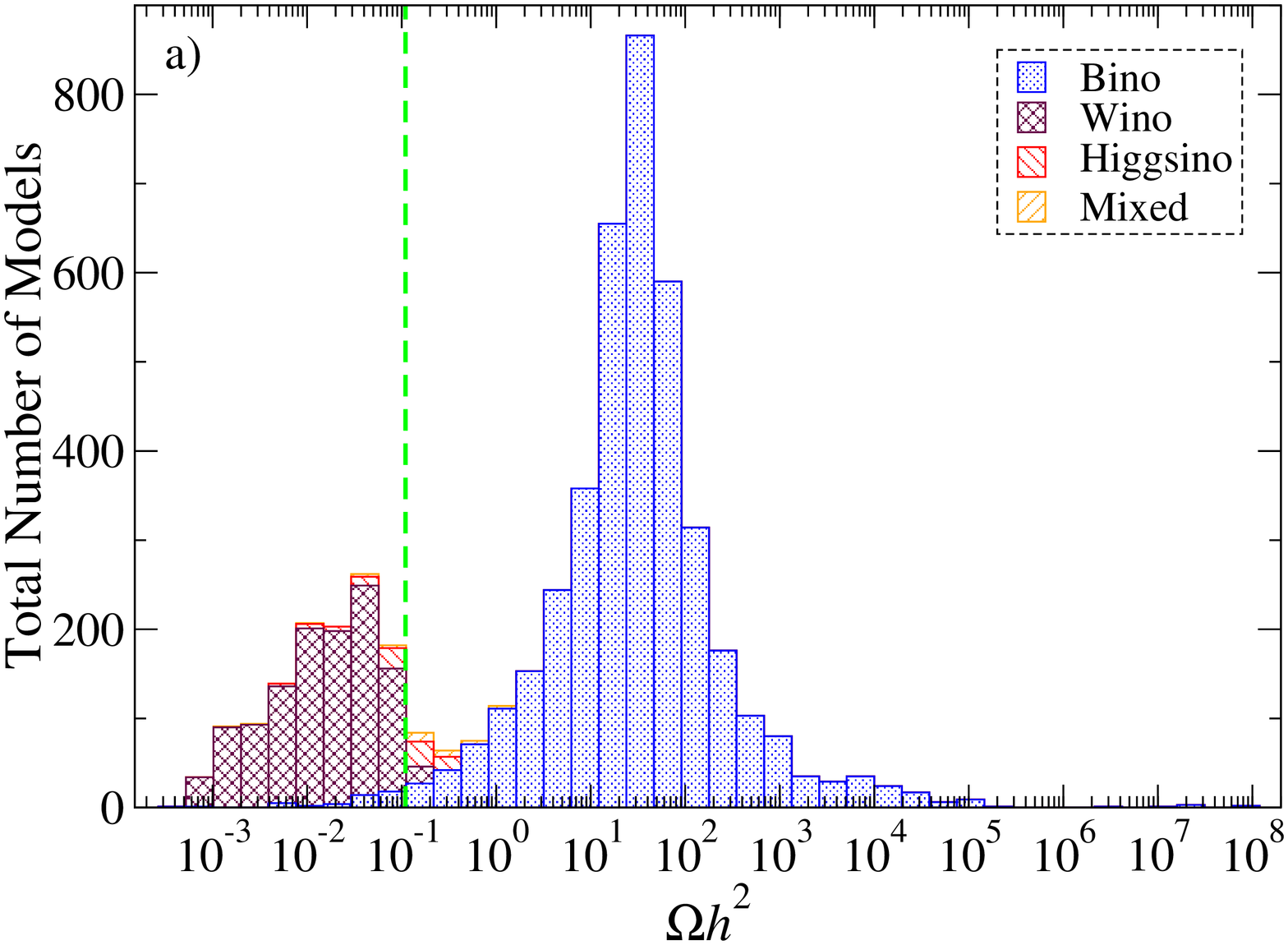}
\includegraphics[angle=-90,width=12cm]{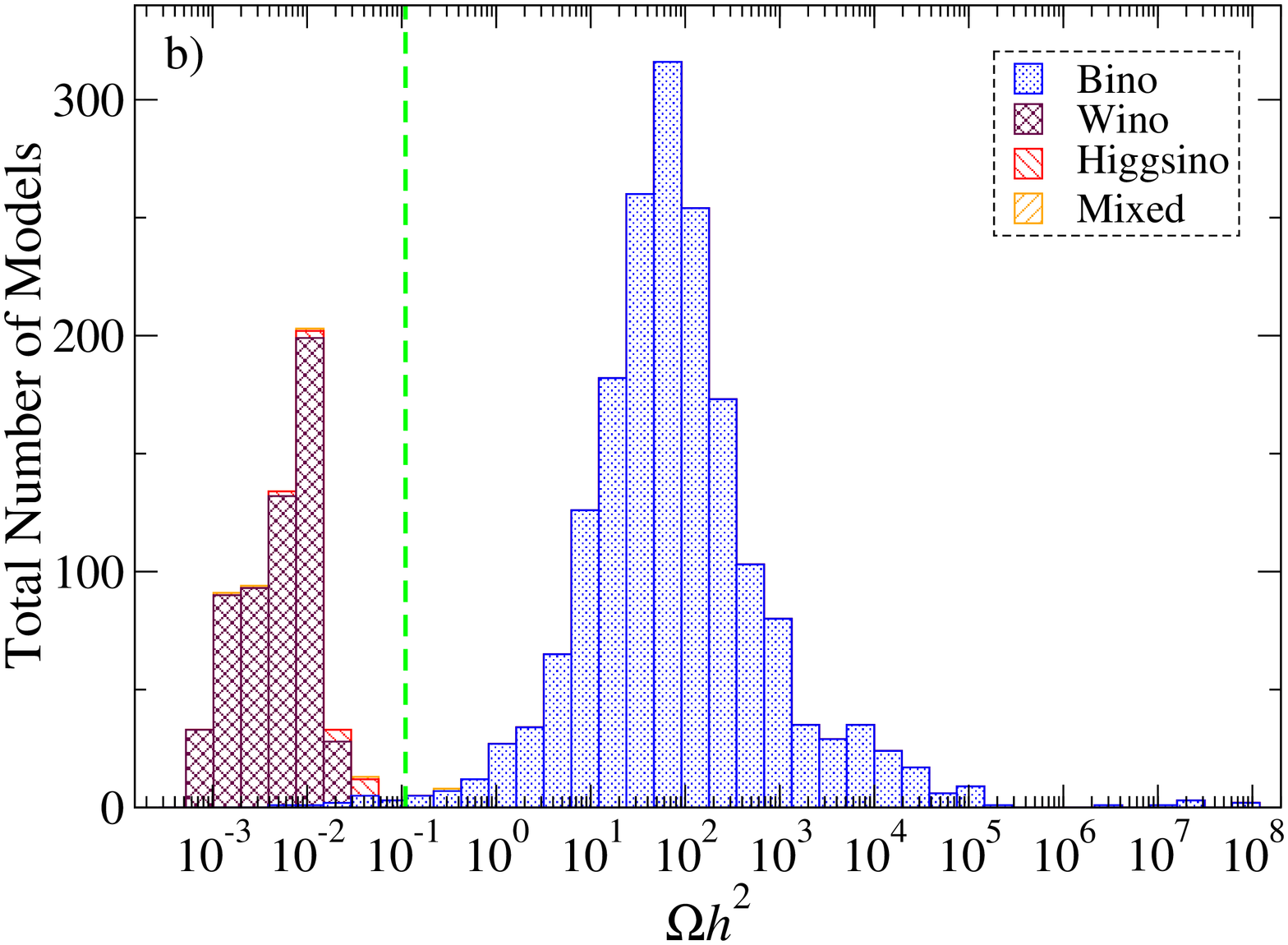}
\caption{Projection of the number of models generated by a linear scan over
SUGRA-19 parameters, versus neutralino relic density $\Omega_{\tz_1}h^2$.
Models with mainly bino, wino, higgsino or a mixture are indicated by the 
various color and symbol choices.
In frame {\it b})., we require only models with $m_{\tz_1}<500$ GeV to avoid
too large of fine-tuning in the SUSY parameters.
}\label{fig:bar1}}

A large number of the wino and higgsino dark matter solutions with
$\Omega_{\tz_1}h^2\sim 0.1$ come from models with very heavy neutralinos: 
$m_{\tz_1}\agt 800$ GeV. If the $\tz_1$ is the LSP, as assumed here, then all
other sparticles are heavier-- and usually much heavier-- than this value,
and will likely lead to solutions with high electroweak fine-tuning\cite{ewft}.
In Fig. \ref{fig:bar1}{\it b})., we plot the number of model solutions from the SUGRA-19 scan
versus $m_{\tz_1}$,  where in addition we require (somewhat arbitrarily) that
$m_{\tz_1}<500$ GeV, so the solutions are not too fine-tuned 
with regard to electroweak symmetry breaking.
In this case, the higgsino and wino LSP models which naturally give
$\Omega_{\tz_1}h^2\sim 0.1$ are all excluded. The maxima of bino-like
solutions has moved up slightly to $\Omega_{\rm bino}h^2\sim 50-100$, while
wino-like solutions peak at $\Omega_{\rm wino}h^2\sim 0.01$. The minimum of the
probability distribution lies very close to the measured 
value $\Omega_{\rm CDM}h^2\sim 0.1$.
With sparticle masses generally at the TeV or below scale,
the measured relic density lies at the {\it least likely value} as predicted by the SUGRA-19
model. In this case, it would be {\it extremely fortuitous}
if the lightest neutralino of SUGRA  theories was in fact the dark matter particle.

\subsection{Log scan over SUGRA-19 parameters}

We have interpreted our linear scan over SUGRA-19 parameter space in terms of
a probability distribution as to likely values of $\Omega_{\tz_1}h^2$ which would be
obtained in supersymmetric models. These results do depend on how we sample
our GUT-scale parameter space. While it is impossible to know what the correct measure is for
GUT-scale SUSY parameters, at least we can compare our linear scan against results using a 
different sampling measure for our parameter space.

In this section, we will instead adopt a logarithmic scan over GUT scale parameters, which favors
lower energy soft SUSY breaking parameters over high energy ones. Specifically,
for a dimensionful parameter $a$ ranging up to $a_{max}$, and a random number $x$ which
is sampled uniformly between 0 and 1, we will generate values of $a$ according to
$a=\exp\left( x\log(a_{max})\right)$.

Applying the log sampling to all dimensionful parameters listed in 
Sec. \ref{sec:method} (but with a linear scan over $\tan\beta$),
and with the same parameter maxima, we re-plot our results in Fig. \ref{fig:scanlog}. 
Here, we see that models with lower values of $m_{\tz_1}$ are sampled much more prominently than 
high values of $m_{\tz_1}$. In the figure, we see that while bino-like models still predict a 
rather high relic density, now there are some bino-like LSP points with relatively small values of $m_{\tz_1}$
and $\Omega_{\tz_1}h^2\sim 0.1$. Meanwhile, the higgsino-like and wino-like bands still persist
at the same location as in Fig. \ref{fig:scan}, albeit with a greater population of points
at low $m_{\tz_1}$. We also obtain some mixed dark matter solutions at low $m_{\tz_1}$.
\FIGURE[t]{
\includegraphics[angle=-90,width=12cm]{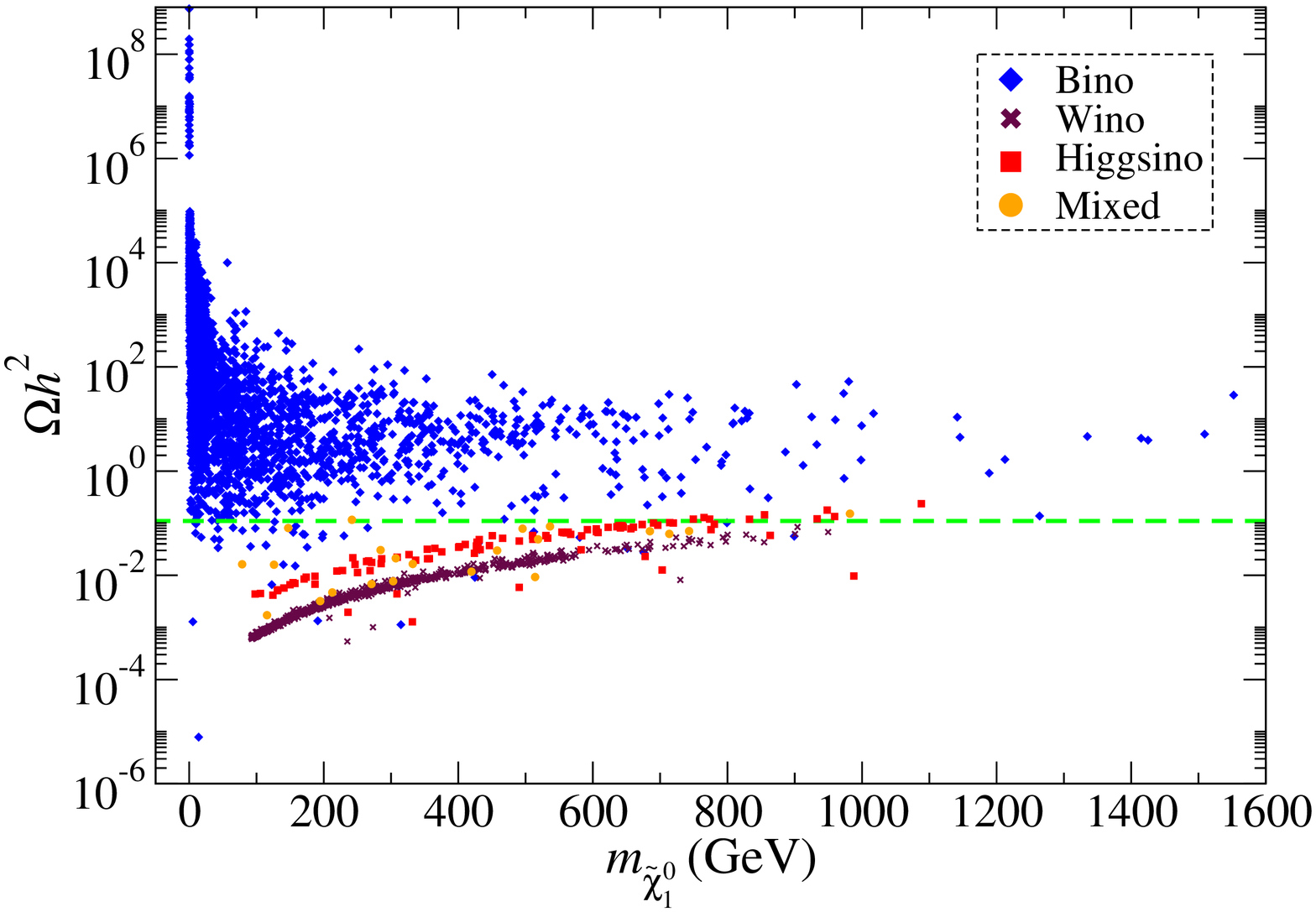}
\caption{Thermal abundance of neutralino cold dark matter from a log scan over
SUGRA-19 model parameter space. We plot versus the neutralino mass. 
Models with mainly bino, wino, higgsino or a mixture are indicated by the 
various color and symbol choices. There are 4276 points in the figure.
}\label{fig:scanlog}}

To compare against the previous distribution in models (Fig. \ref{fig:bar1}), we again project the
models as a histogram on the $\Omega_{\tz_1}h^2$ axis in Fig. \ref{fig:bar1log}{\it a}).
While the minimum in the probability distribution still lies nearly at the measured CDM abundance, 
the minimum has filled in somewhat, especially with a mixture of bino-like and higgsino-like solutions.
\FIGURE[t]{
\includegraphics[angle=-90,width=12cm]{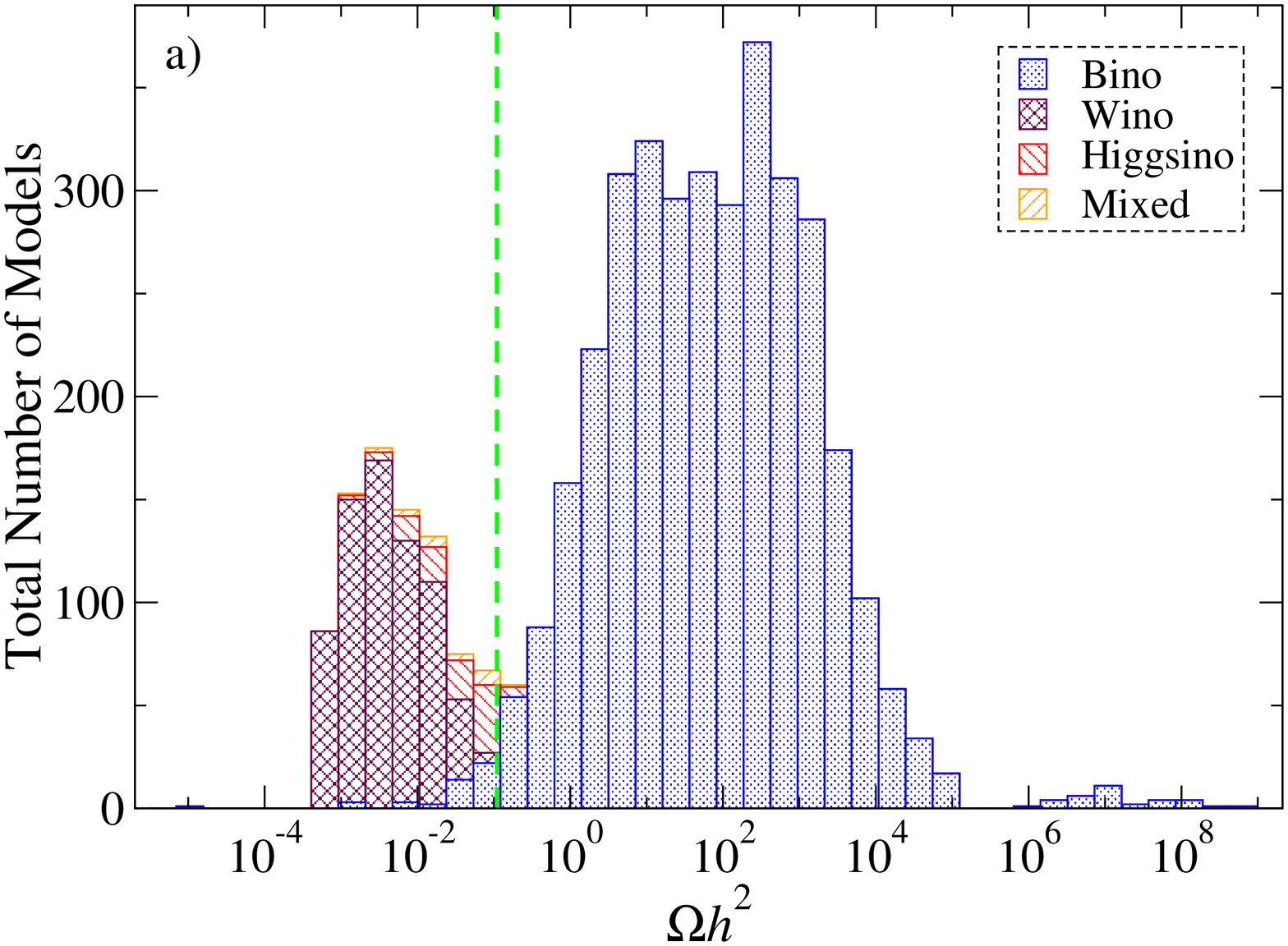}
\includegraphics[angle=-90,width=12cm]{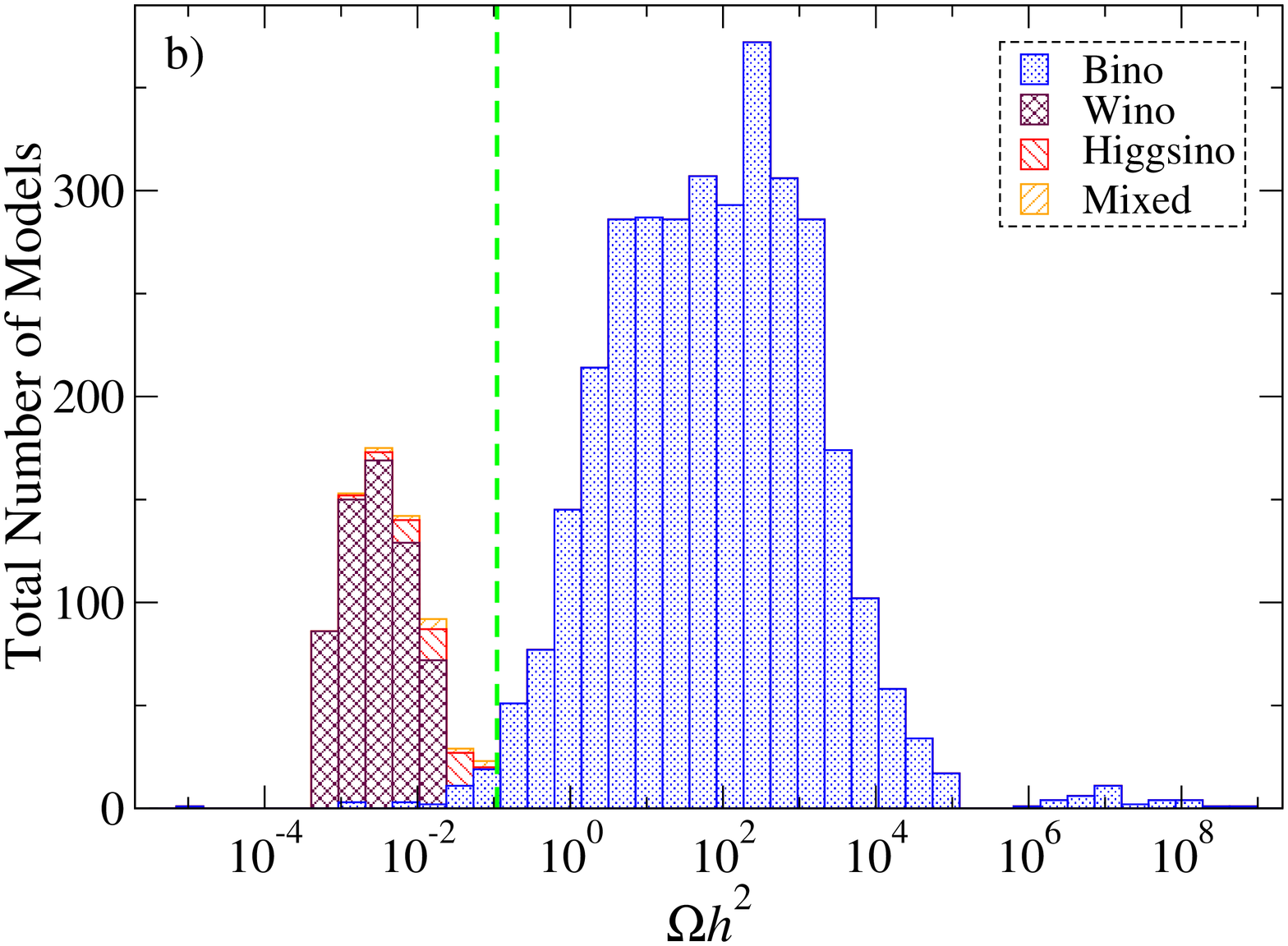}
\caption{Projection of the number of models generated by a log scan over
SUGRA-19 parameter space, versus neutralino relic density $\Omega_{\tz_1}h^2$.
In frame {\it b})., we require only models with $m_{\tz_1}<500$ GeV to avoid
too large of fine-tuning in the SUSY parameters.
Models with mainly bino, wino, higgsino or a mixture are indicated by the 
various color and symbol choices.
}\label{fig:bar1log}}

For comparison with Fig. \ref{fig:bar1}{\it b})., 
we apply the requirement of $m_{\tz_1}<500$ GeV and gain
the results shown in Fig. \ref{fig:bar1log}{\it b}). 
In this case, the minimum probability has migrated to 
slightly lower values, as the higgsino-like and wino-like solutions with 
$\Omega_{\tz_1}h^2\sim 0.1$ require $m_{\tz_1}\agt 800$ GeV, and have been rejected.
We still have a few bino-like models persisting with $\Omega_{\tz_1}h^2\sim 0.1$, although the
probability is again quite small (but not as small as in the linear sampling case).

\subsection{Comparison of SUGRA-19 results to the mSUGRA model}
\label{sec:sugra}

In this section, for comparison purposes, we present similar results if
we restrict our scan to the mSUGRA model with the 
well-known parameter space
\be
m_0,\ m_{1/2},\ A_0,\ \tan\beta ,\ sign(\mu ),
\ee
with $m_t=173.1$ GeV as before.
We sample $m_0:0-6$ TeV, $m_{1/2}:0-2$ TeV, $A_0:-3.5\to 3.5$ TeV and $\tan\beta :2\to 60$.
This expands upon the treatment presented in Ref. \cite{bulk}, since it provides a general
scan over mSUGRA parameters.
We only present results of the linear scan for this case.
The results of our scan are shown in Fig. \ref{fig:sugscan}, in the $\Omega_{\tz_1}h^2\ vs.\ m_{\tz_1}$ plane.
We again see that bino-like neutralino solutions populate the high $\Omega_{\tz_1}h^2$ 
region, while higgsino-like and mixed bino-higgsino dark matter solutions (arising from the focus point region) 
populate the 
range from $\Omega_{\tz_1}h^2\sim 0.01-0.3$. The nearly pure higgsino solutions are almost always 
lower than the measured DM abundance. The wino-like CDM solutions have vanished, since in this case,
with gaugino mass unification at $M_{GUT}$, we always get gaugino masses $M_1\ll M_2$ at the weak scale.
\FIGURE[t]{
\includegraphics[angle=-90,width=12cm]{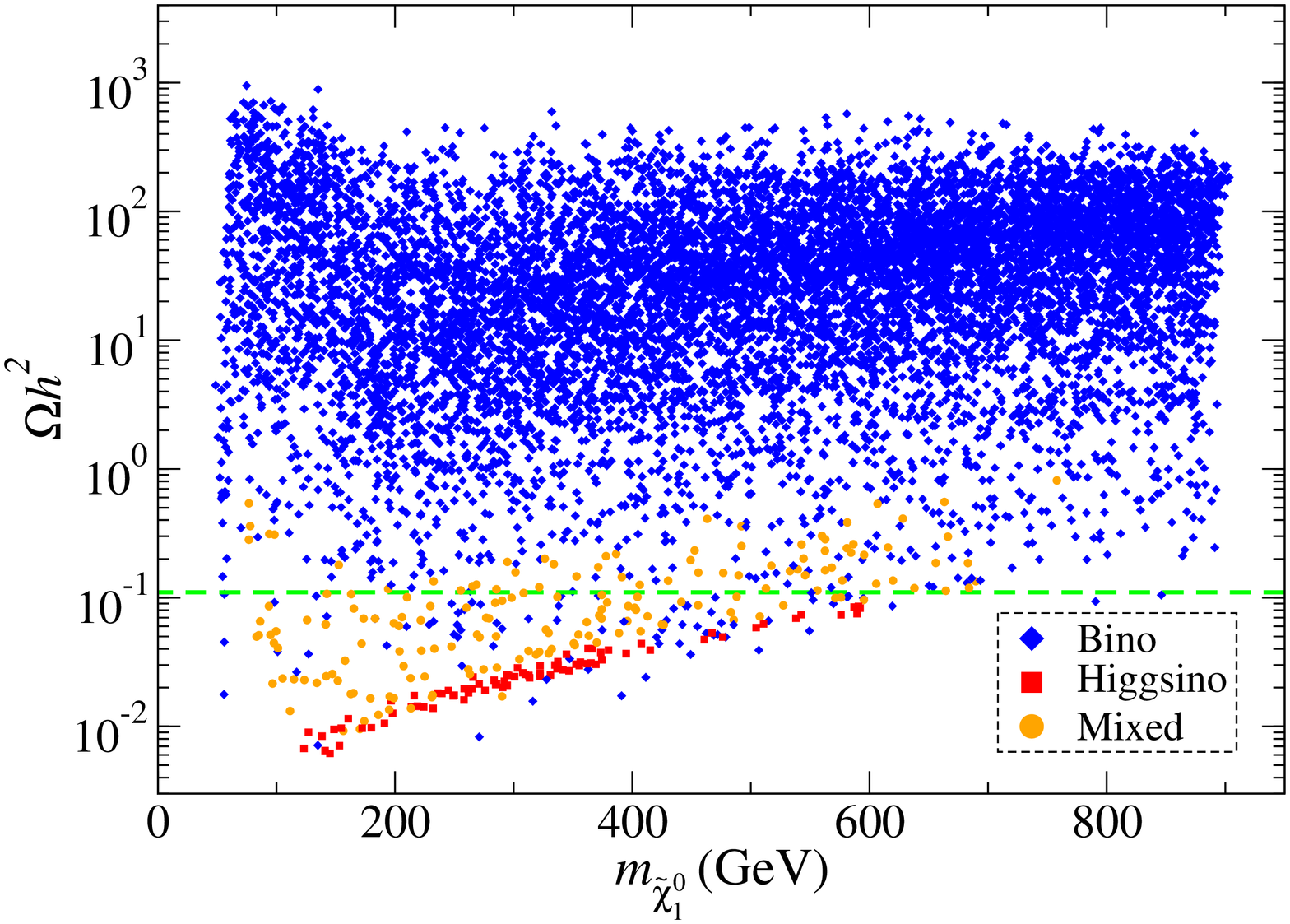}
\caption{Thermal abundance of neutralino cold dark matter from a linear scan over
the mSUGRA model parameter space. We plot versus the neutralino mass. 
Models with mainly bino, wino, higgsino or a mixture are indicated by the 
various color and symbol choices. There are 10,025 points in the figure.
}\label{fig:sugscan}}

In Fig. \ref{fig:sugbar1}{\it a})., we project the solutions onto the $\Omega_{\tz_1}h^2$
axis. The prevalence of bino-like DM solutions is clear, with a maximum around
$\Omega_{\tz_1}h^2\sim 30-50$. A tail drops off gradually at low values
of $\Omega_{\tz_1}h^2$, although the tail flattens for a while when it is augmented by the
appearance of higgsino-like and mixed bino-higgsino DM solutions near the measured value
of $\Omega_{\tz_1}h^2\sim 0.1$.
\FIGURE[t]{
\includegraphics[angle=-90,width=12cm]{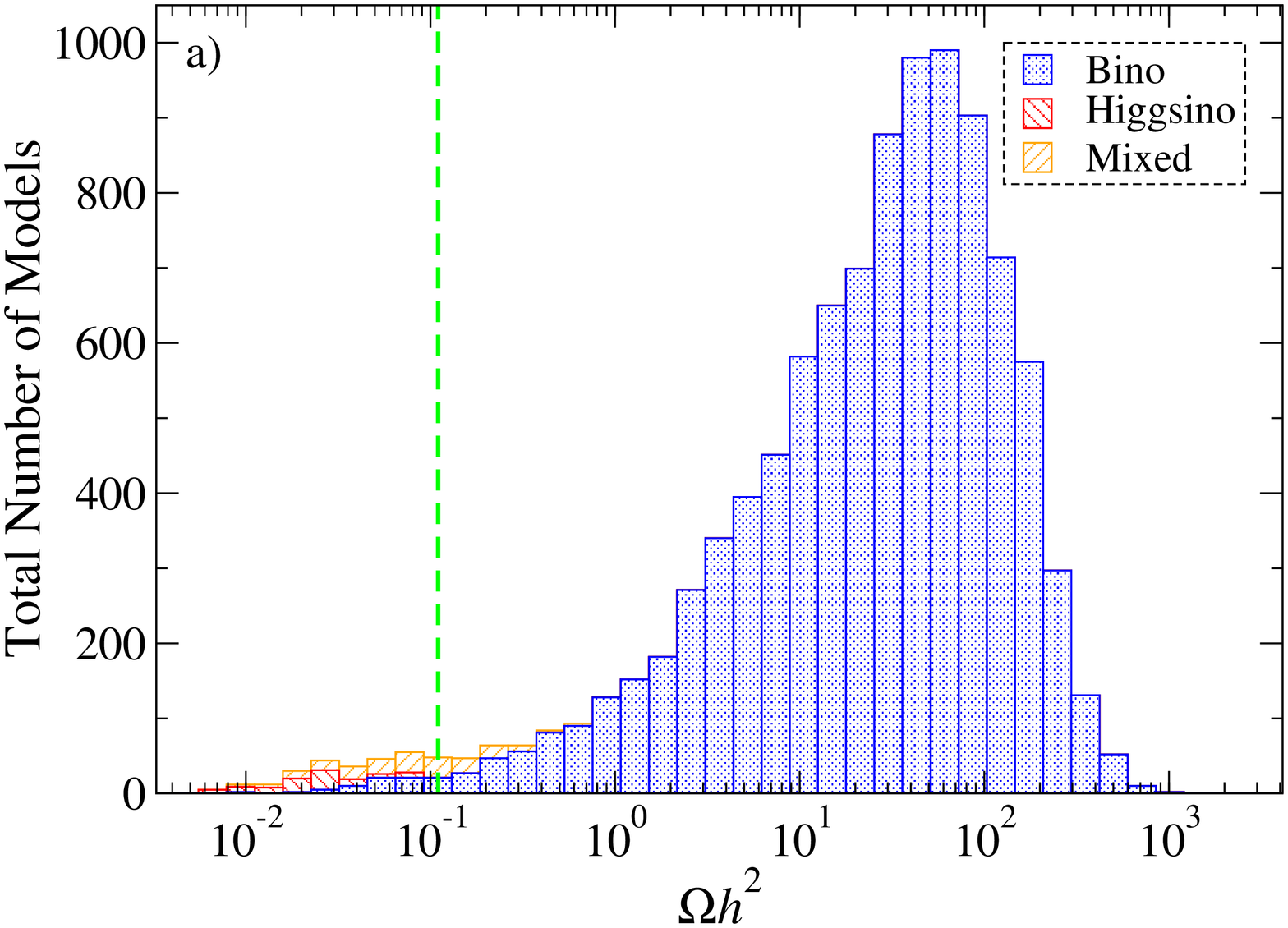}
\includegraphics[angle=-90,width=12cm]{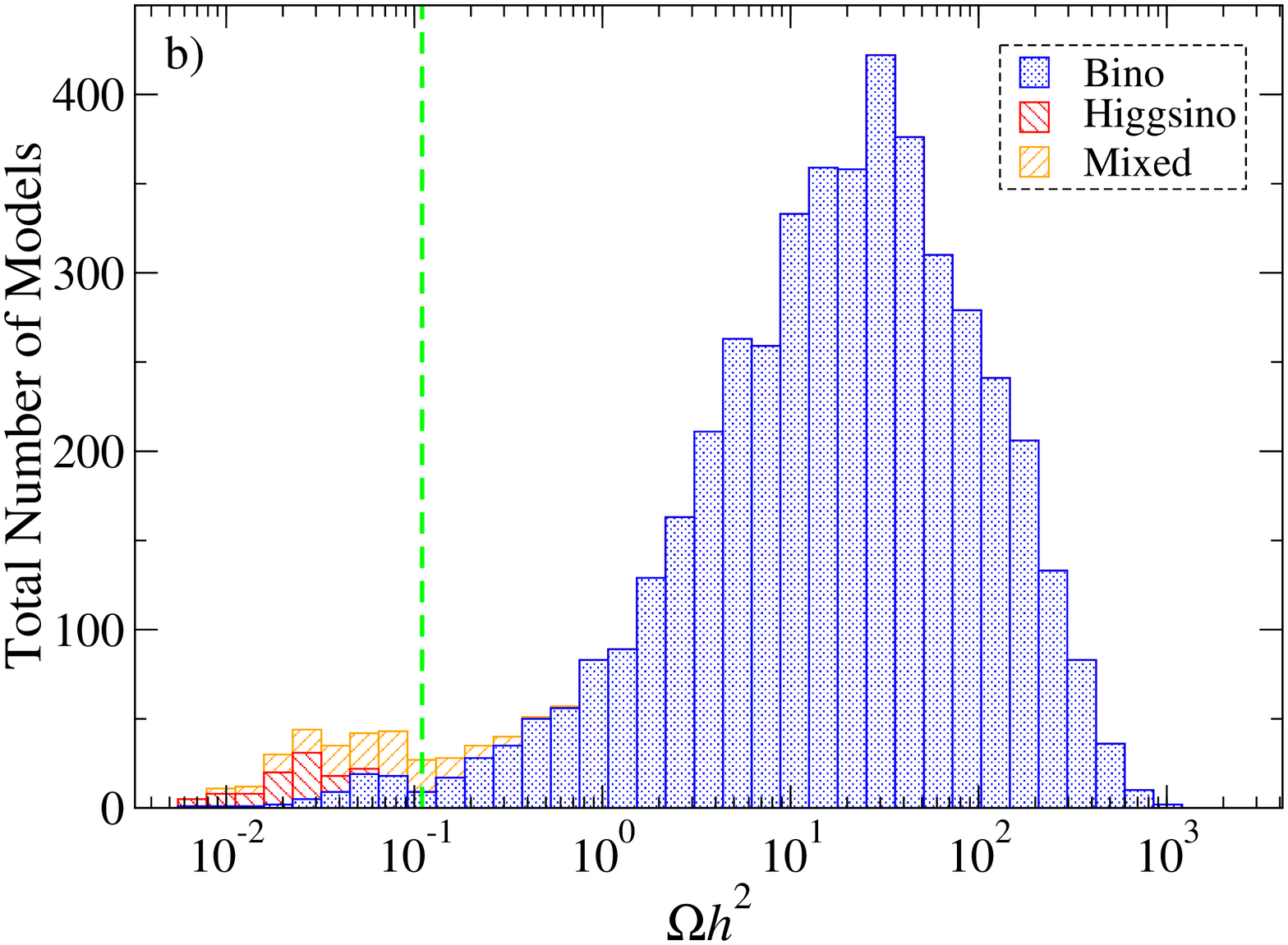}
\caption{Projection of the number of models generated by a linear scan over
mSUGRA parameters, versus neutralino relic density $\Omega_{\tz_1}h^2$.
In frame {\it b})., we require only models with $m_{\tz_1}<500$ GeV to avoid
too large of fine-tuning in the SUSY parameters.
Models with mainly bino, wino, higgsino or a mixture are indicated by the 
various color and symbol choices.
}\label{fig:sugbar1}}

If we apply the rough electroweak naturalness requirement that $m_{\tz_1}<500$ GeV, we arrive at
the results in Fig. \ref{fig:sugbar1}{\it b}).
In this case, many of the higgsino-like solutions with $\Omega_{\tz_1}h^2\sim 0.1$ are now excluded, 
and the measured relic density lies again at a slight dip between the large bino-like maximum
and the softer mixed bino-higgsino maximum around $\Omega_{\tz_1}h^2\sim 0.03$.
By comparing Fig. \ref{fig:sugbar1}{\it b}). for the mSUGRA model against Fig. \ref{fig:bar1}{\it b}). 
for SUGRA-19,
we conclude that it is actually easier to generate the measured abundance of CDM 
in mSUGRA than in the more expansive SUGRA-19 parameter space.

\section{Scan over PQMSSM with mixed axion/axino cold dark matter}
\label{sec:axino}

In this section, we assume the MSSM augmented by the Peccei-Quinn mechanism (PQMSSM)
is the correct effective field theory below the scale $T_R$. Thus, the MSSM is supplemented by
an axion supermultiplet  containing the spin-zero and $R$-parity even  axion $a$ and saxion $s$ fields, 
and the $R$-odd spin-$1\over 2$ axino field $\ta$. 
We assume the axino is the LSP, and one of the constituents of dark matter.
We do not assume a specific axion
model, {\it e.g.} KSVZ\cite{ksvz} or DFSZ\cite{dfsz}, nor do we assume a specific 
SUSY breaking mechanism to give the axino mass. We parametrize the low energy
axion/axino physics by adopting as free parameters the PQ breaking scale $f_a/N$ (where 
$N$ is the model-dependent color anomaly), the axino mass $m_{\ta}$ (for which $m_{\ta}<m_{\tz_1}$) 
the re-heat temperature $T_R$ (which is related to unknown inflaton parameters) and the
axion field initial mis-alignment angle $\theta_i$.

We will assume a re-heat temperature $T_R$ bounded by $min[ 10^{10}\ {\rm GeV},T_{dcp}]$,
where $T_{dcp}\sim 10^{11}\left(\frac{ f_a/N}{10^{12}\ {\rm GeV}}\right)^2$ GeV, 
so as to be consistent with results from Sec. \ref{sec:wimp}, which avoided 
overclosure limits from excessive gravitino production in the early universe\cite{gravproblem}.
This limit also avoids the RTW bound\cite{rtw} $T_R\agt T_{\rm dcp}$
for which axinos would be in thermal equilibrium in the early universe.\footnote{
In this case, if $T_R>T_{dcp}$, then $m_{\ta}<.2$ keV to avoid overclosure due to 
excessive axino production\cite{rtw,steffen}. 
Such light axinos would constitute {\it hot} dark matter.}
\footnote{If $T_R\agt f_a$, then PQ symmetry is restored after inflation, 
and when the Universe cools, PQ symmetry is re-broken, and various domains within a Hubble
volume may have different values of mis-alignment angle $\theta_i$. Our upper limit
on $T_R$ excludes this scenario (Scenario I of Ref. \cite{vg}).}
We will also assume $m_{\ta}\agt 100$ keV; for lighter axino masses, the
thermally produced axinos (see below) would likely constitute warm, rather than cold, 
dark matter\cite{jlm}. 
Thus, we supplement the SUGRA-19 model by an axion/axino supermultiplet along 
with the additional parameters
\bi
\item $f_a/N:\ 10^9\to 10^{16}$ GeV,
\item $m_{\ta}:\ 100\ {\rm keV}\to m_{\tz_1}$,
\item $T_R: max[0.01\ {\rm GeV},m_{\tz_1}/20]\to min[10^{10}\ {\rm GeV},T_{dcp}]$, 
\item $\theta_i:-\pi \to\pi$.
\ei
The lower bound on $f_a/N$ comes from bounds on red giant cooling and supernova 1987a. 
The upper bound on $f_a/N$ we take near the GUT scale. The more common bound
$f_a/N\alt 10^{12}$ GeV comes from assuming $\theta_i\sim 1$. The lower bound on $T_R$ 
comes from requiring $T_R\agt 10$ MeV so that standard BBN can occur\cite{bbn}, and requiring
neutralino production at temperatures above neutralino freeze-out.

\subsection{Mixed axion/axino relic density}

In this section, we briefly review dark matter production in the PQMSSM, with an axino as LSP.
Since we assume a value $T_R\alt 10^{10}$ GeV, then axinos should never be in thermal
equilibrium in the early universe\cite{rtw}. 
However, thermal production of neutralinos (now assumed to be the NLSP) should proceed as usual, 
except that now each neutralino will decay via $\tz_1\to\ta\gamma$. Since the $\tz_1$ lifetime is
of order $0.1$ sec, it is assumed to be BBN-safe.
This non-thermal contribution to the axino CDM abundance is given by
\be
\Omega_{\ta}^{NTP}=\frac{m_{\ta}}{m_{\tz_1}}\Omega_{\tz_1}h^2
\label{eq:Oh2_NTP}
\ee
since the axino number density from this source is the same as the thermally produced neutralino
number abundance.

While the axinos are assumed never in thermal equilibrium in the early universe, 
nevertheless, they can still be produced thermally
via radiation off of particles that are in thermal equilibrium, {\it i.e.} the quarks, leptons, gluons,
SUSY particles {\it etc.}. This thermal axino production mechanism has been calculated 
in Ref's \cite{ckkr}, \cite{steffen} and \cite{strumia}. 
The latter calculation includes some interaction terms neglected in previous calculations; 
they find
\be
\Omega_{\ta}^{TP}h^2\simeq 1.24 g_s^4 F(g_s)\frac{m_{\ta}}{{\rm GeV}}
\frac{T_R}{10^4\ {\rm GeV}}\left(\frac{10^{11}\ {\rm GeV}}{f_a/N}\right)^2
\ \ \ F(g_s)\simeq 20g_s^2\ln\frac{3}{g_s}
\label{eq:Oh2_TP}
\ee
where $g_s$ is the strong coupling evaluated at $Q=T_R$ ({\it e.g.} $g_s(T_R=10^6\ {\rm GeV})=0.932$, 
as given by Isajet 2-loop $g_s$ evolution in mSUGRA).

Finally, relic {\it axions} will be produced via the vacuum mis-alignment mechanism.
In this case, we use\cite{absik,vg} 
\be
\Omega_a h^2 = {1\over 4} f(\theta_i) \theta_i^2\left(
\frac{f_a/N}{10^{12}\ {\rm GeV}}\right)^{7/6},
\ee
where $f(\theta_i)$ is the anharmonicity factor. We adopt a recent parametrization of 
$f(\theta_i )$ from Visinelli and Gondolo\cite{vg} given by
\be
f(\theta_i)=\left[\ln\left(\frac{e}{1-\theta_i^2/\pi^2}\right)\right]^{7/6}.
\ee
The function $f(\theta_i)\to 1$ as $\theta_i\to 0$ and $f(\theta_i)\to\infty$ as $\theta_i\to\pm\pi$.

The total dark matter abundance in the PQMSSM with an axino LSP is then given by the sum of the three constituents:
\be
\Omega_{a\ta}h^2=\Omega_{\ta}^{NTP}h^2+\Omega_{\ta}^{TP}h^2+\Omega_{a}h^2 .
\ee

\subsection{Results of scan over SUGRA-19 with mixed $a\ta$ CDM}

Our first results from the linear scan over SUGRA-19 model parameters augmented by the
four PQMSSM parameters are shown in Fig. \ref{fig:scan_fa}, where we plot
models generated in the $\Omega_{a\ta}h^2\ vs.\ f_a/N$ plane. 
Models with dominant
axion CDM are in red, while models with dominant axino CDM are in blue.
The measured CDM abundance is denoted by the dashed green line.
The densely populated diagonal band which increases as $f_a/N$ increases is due to 
mainly axion CDM with $\theta_i\sim 1$. The diffuse dotted region at low $f_a/N$, which extends up to very large
values of $\Omega_{a\ta}h^2\sim 10^{9}$ is due to thermally produced axinos. For thermally
produced axinos, the axino-matter coupling is inversley proportional to $f_a/N$, and so the 
coupling is large at low $f_a/N$ values, leading to the large thermal abundance. As $f_a/N$ increases,
the axino-matter coupling is suppressed, while the magnitude of the axion vacuum energy increases, 
leading to the case of mixed axion/axino CDM which is mainly axions.
\FIGURE[t]{
\includegraphics[angle=0,width=12cm]{faomgh2.eps}
\caption{SUGRA-19 parameter space points from a linear scan
where we assume mixed axion/axino CDM. There are 4217 points in the figure.
}\label{fig:scan_fa}}

In Fig. \ref{fig:scan_tr}, we plot models in the $\Omega_{a\ta}h^2\ vs.\ T_R$ plane. 
In this case, we see the models
are spread over a wide range of $T_R$ values. Large values of $T_R$ give rise to a large rate for
thermally produced axinos, and so $\Omega_{a\ta}$ increases somewhat with increasing $T_R$. In fact, it
becomes difficult, but not impossible, to gain a value of $\Omega_{a\ta}h^2\sim 0.1$ while maintaining
$T_R\agt 10^6$ GeV, as needed for baryogenesis processes such as non-thermal leptogenesis\cite{ntlepto}.
Since the axion abundance doesn't depend on $T_R$, the low $T_R$ points are dominanted by mainly axion
CDM models.
\FIGURE[t]{
\includegraphics[angle=0,width=12cm]{tromgh2.eps}
\caption{Scan over the PQMSSM  parameter space plotted in the
$\Omega_{a\ta}h^2\ vs.\ T_R$ plane
}\label{fig:scan_tr}}

In Fig. \ref{fig:scan_max}, we plot models in the $\Omega_{a\ta}h^2\ vs.\ m_{\ta}$ plane. 
Here, we see the models are spread rather uniformly over many decades of $m_{\ta}$, although
the tendency is for $\Omega_{a\ta}h^2$ to increase as $m_{\ta}$ increases. This is partly due to 
the fact that each relic axino is more massive.
\FIGURE[t]{
\includegraphics[angle=0,width=12cm]{axinoomgh2.eps}
\caption{SUGRA-19 parameter space points from a linear scan
where we assume mixed axion/axino CDM.
}\label{fig:scan_max}}

To compare against results from our thermal neutralino calculations of Sec. \ref{sec:wimp}, we 
project all the models onto the $\Omega_{a\ta}h^2$ axis, and present our results as a histogram 
in the number of models generated. We see from frame {\it a}). of Fig. \ref{fig:axhist} that the models are
spread over many decades of $\Omega_{a\ta}h^2$. The measured abundance-- denoted by the
green dashed line-- lies on the rising shoulder of the distribution, which in fact peaks
around $\Omega_{a\ta}h^2\sim 10^2$. If we require in addition $m_{\tz_1}<500$ GeV as in Fig. \ref{fig:bar1}{\it b}).
as a rough EW fine-tuning requirement, then we obtain the results of frame {\it b}). 
Qualitatively, the two plots are similar, and the measured abundance still lies on
the increasing shoulder of the distribution, which again peaks around $\Omega_{a\ta}h^2\sim 100$.
\FIGURE[t]{
\includegraphics[angle=-90,width=12cm]{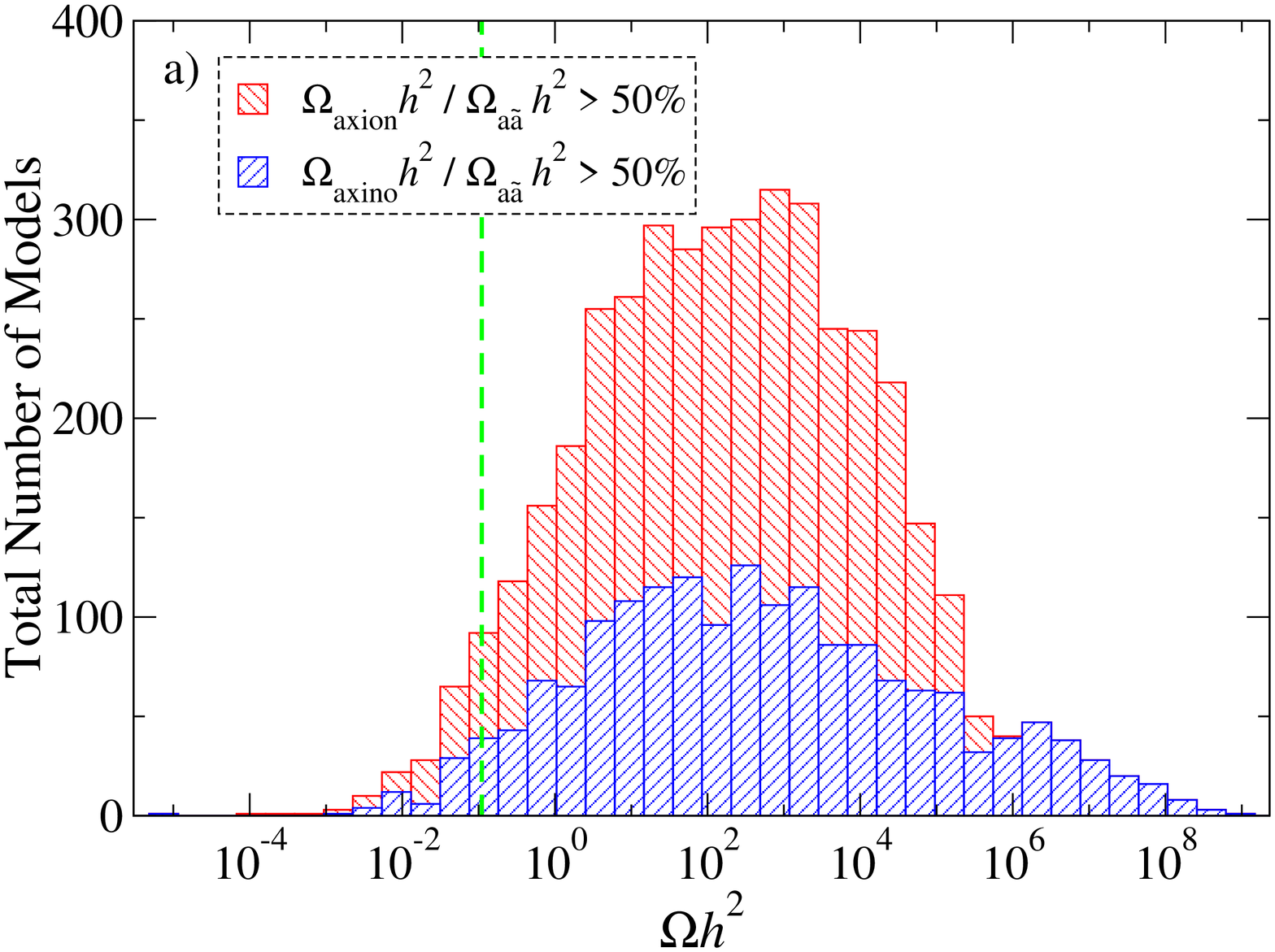}
\includegraphics[angle=-90,width=12cm]{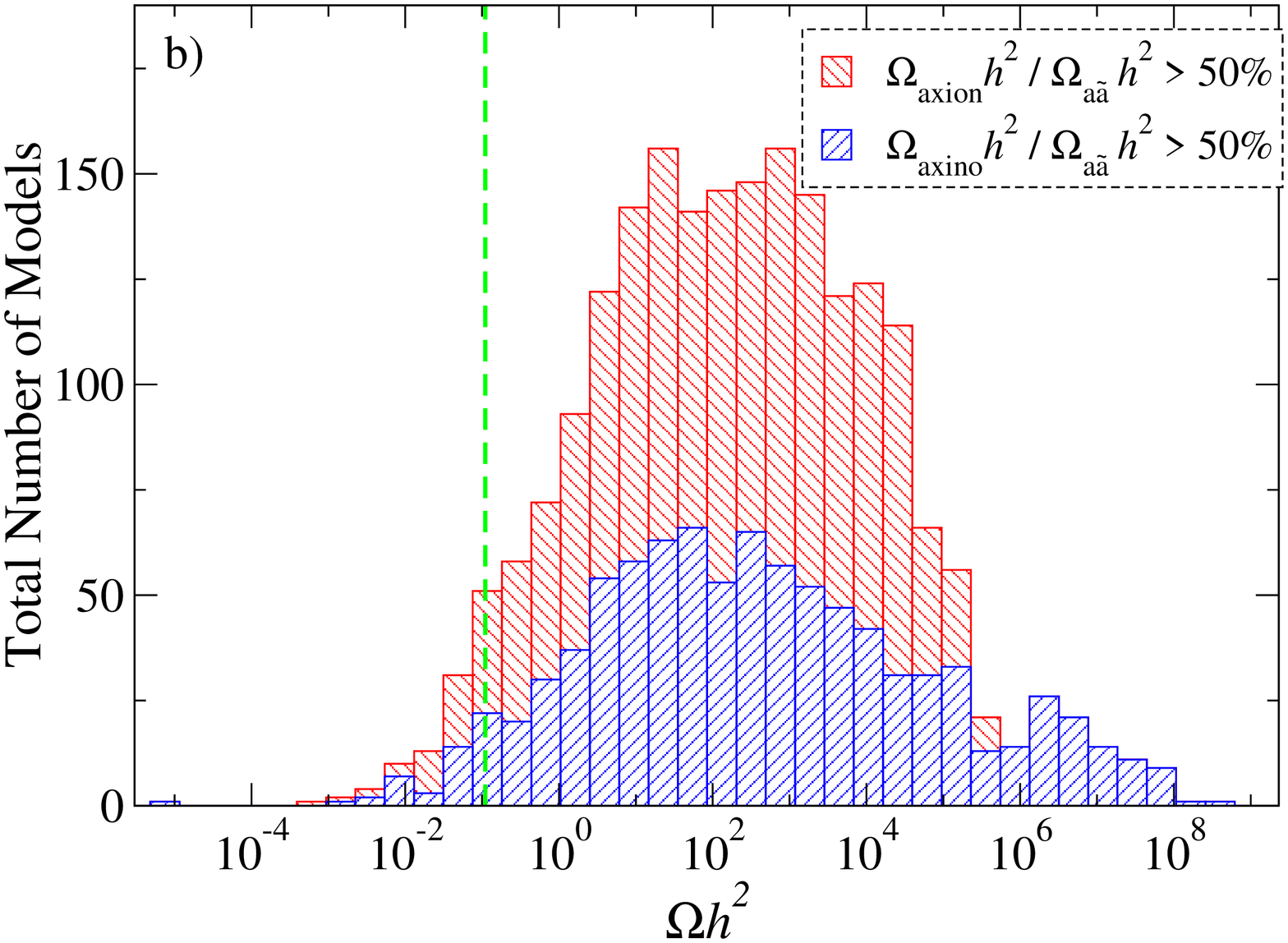}
\caption{Number of models versus $\Omega_{a\ta}h^2$ in the 
scan over the PQMSSM  parameter space. Frame {\it a}). contains all
models. Frame {\it b}). requires $m_{\tz_1}<500$ GeV, 
as a rough constraint due to electroweak fine-tuning.
}\label{fig:axhist}}

\subsection{PQMSSM parameters which are preferred by the measured CDM abundance}

As a final exercise, it may be instructive to see which values of PQMSSM parameters
are preferred by the measured abundance of CDM in the universe. To this end, we
will require $\Omega_{a\ta}h^2$ to lie between 0.05 and 0.2: {\it i.e.} within a factor of two
of the measured abundance. Then we can project our random scans on the parameter axis.

In Fig. \ref{fig:slice_fa}, we project the PQMSSM models onto the $f_a/N$ plane. In this case, 
we see that the measured CDM abundance prefers a value of $f_a/N\sim (1-2)\times 10^{11}$ GeV, 
with broad tails extending to higher and lower values. This value of $f_a/N$ corresponds to that 
which is needed for PQMSSM models with mainly {\it axion} CDM and initial mis-alignment angle of 
order 1-2.
\FIGURE[t]{
\includegraphics[angle=-90,width=12cm]{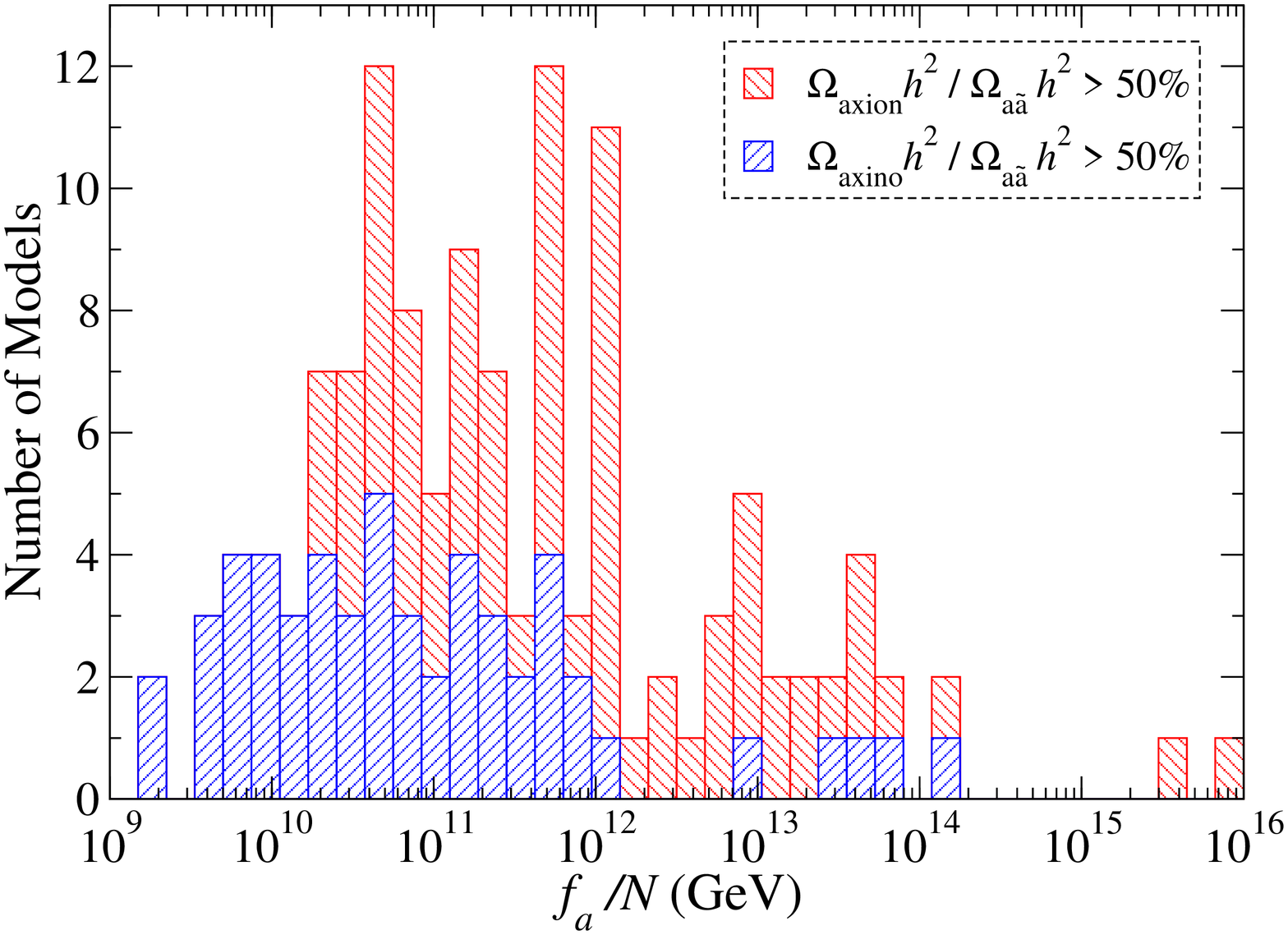}
\caption{SUGRA-19 parameter space points from a linear scan
where we assume mixed axion/axino CDM.
}\label{fig:slice_fa}}

In Fig. \ref{fig:slice_ma}, we show PQMSSM models with the measured abundance versus 
axino mass $m_{\ta}$. Here, the value of $m_{\ta}$ is spread over many decades, with some favoritism for
small values of $m_{\ta}\sim$ MeV range. These again correspond to the models with
mainly axion CDM.
\FIGURE[t]{
\includegraphics[angle=-90,width=12cm]{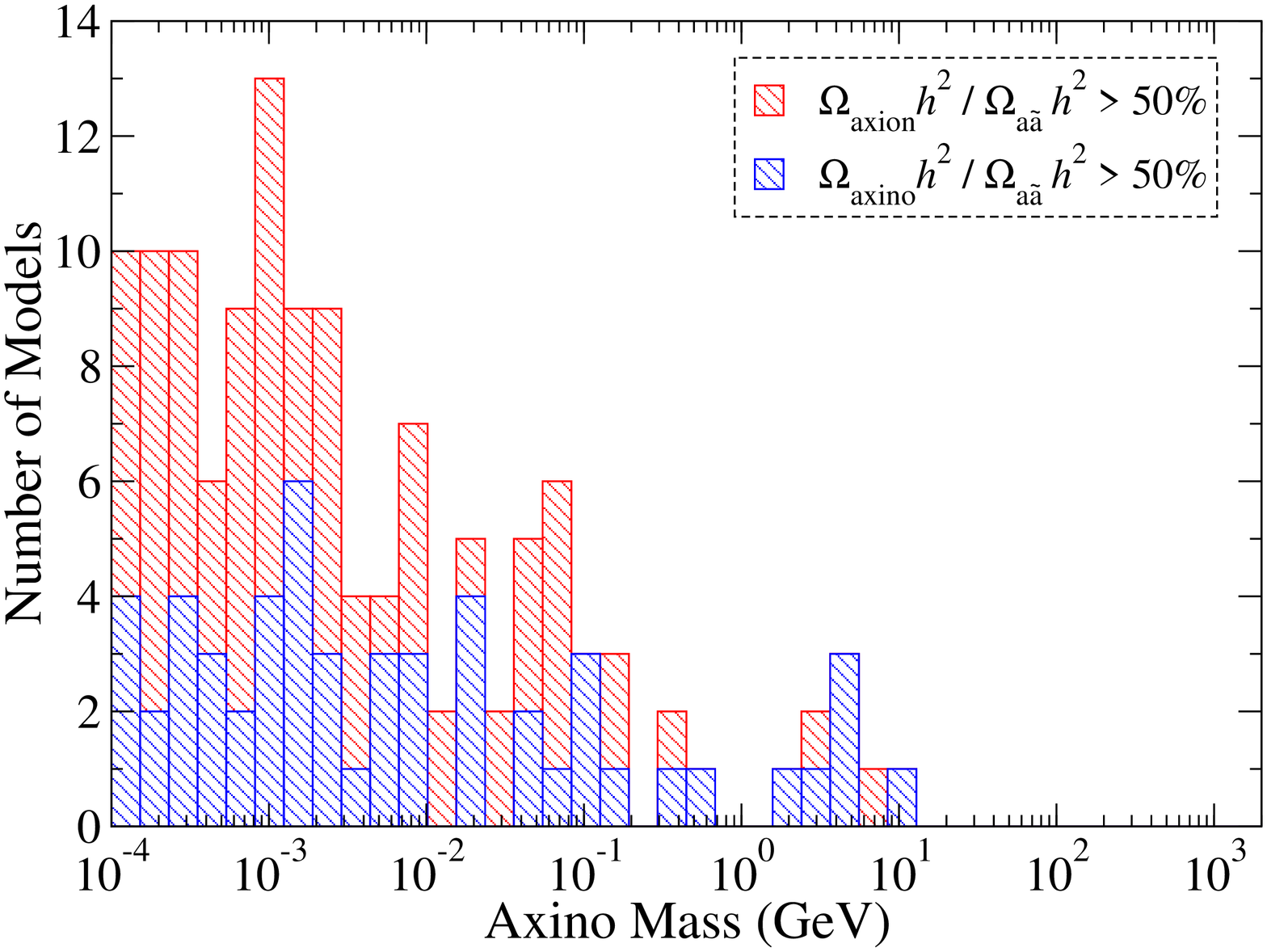}
\caption{SUGRA-19 parameter space points from a linear scan
where we assume mixed axion/axino CDM.
}\label{fig:slice_ma}}

In Fig. \ref{fig:slice_tr}, we show PQMSSM models with the measured abundance versus 
re-heat temperature $T_R$. The $T_R$ values are spread out over many decades. There is some 
preference for low values of $T_R$ in the GeV range, since these models suppress excess
thermal axino production while allowing for vacuum mis-alignment production of 
relic axions, which don't depend on $T_R$. The high $T_R$ tail does extend up to
values of $T_R$ as high as $10^6-10^7$ GeV, which can allow for baryogenesis via non-thermal leptogenesis.
\FIGURE[t]{
\includegraphics[angle=-90,width=12cm]{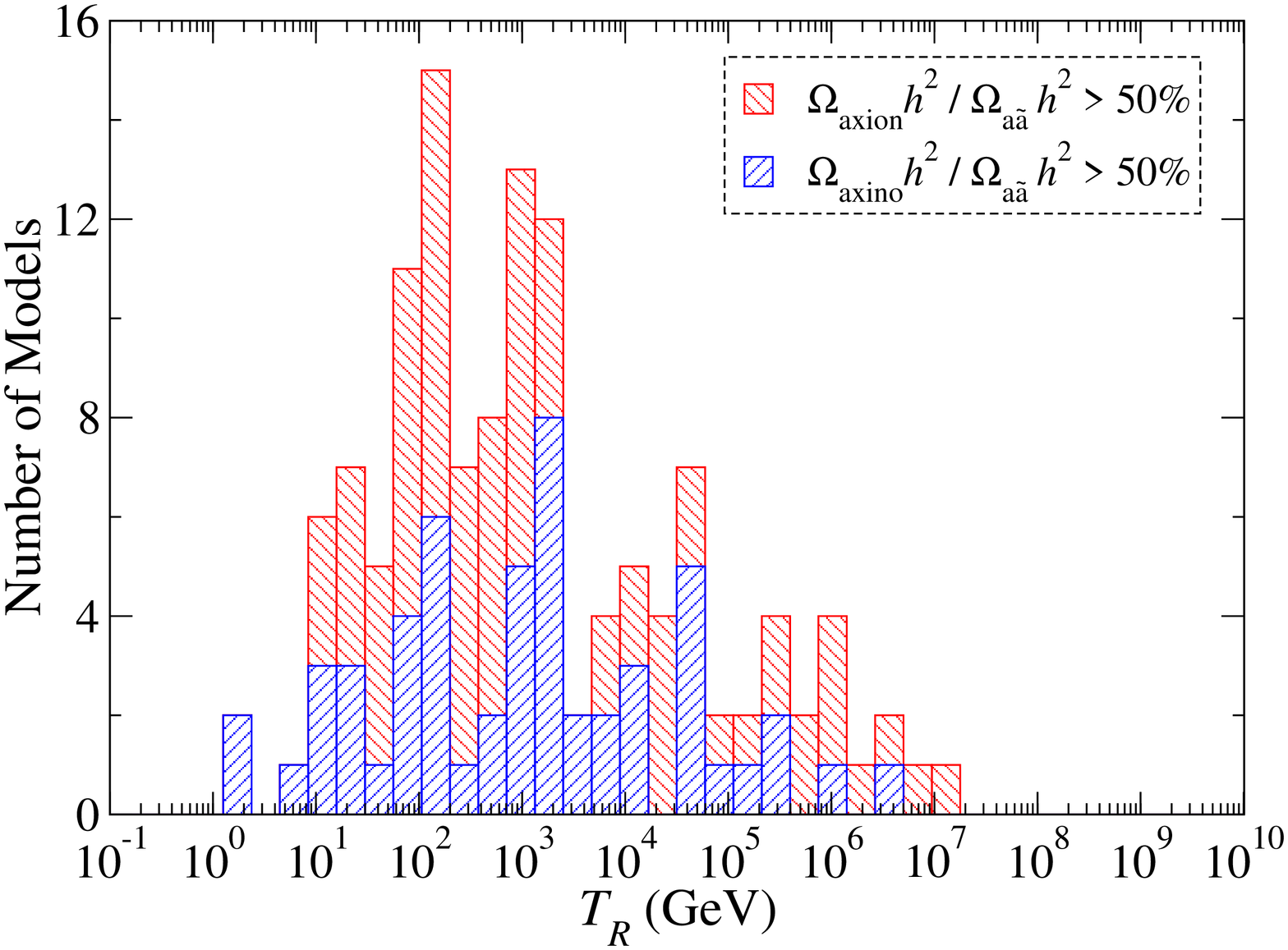}
\caption{SUGRA-19 parameter space points from a linear scan
where we assume mixed axion/axino CDM.
}\label{fig:slice_tr}}
%

%%%%%%%%%%%%%%%%%%%%%%%%%%%%%%%%%%%%%%%%%%%%%%%%%
\section{Summary and conclusions}
\label{sec:conclude}
%%%%%%%%%%%%%%%%%%%%%%%%%%%%%%%%%%%%%%%%%%%%%%%%%

In this paper, we have reported on a calculation of the thermal 
abundance of relic neutralinos in the SUGRA-19 model with 19 free parameters, 
where all dimensional parameters are stipulated at the GUT scale, as suggested by
gauge coupling unification, and SUSY GUTs\cite{drw}.
We find in this rather general framework that bino-like neutralinos populate
the high $\Omega_{\tz_1}h^2\sim 10^1-10^3$ region, while wino and higgsino-like
neutralinos tend to populate the $\Omega_{\tz_1}h^2\sim 10^{-3}-10^{-2}$ region.
The measured CDM abundance sits at a deep minimum between these two favored values,
especially if one requires rather light sparticles with $m_{\tz_1}\alt 500$ GeV, as suggested
by electroweak fine-tuning arguments. In this region, a ``well-tempered neutralino'' is needed;
however, these WTNs with the correct relic abundance are very difficult to generate using GUT scale
SUSY breaking parameters. In this case, we conclude that it would be a near miracle if SUSY neutralinos
were to constitute the measured CDM in the universe. This is the opposite conclusion to that suggested
by the ``WIMP miracle'', where it is asserted that WIMPs have exactly the right properties 
to constitute thermal relics of the Big Bang. In the case of SUSY WIMPs, this does not appear to be so.

Our results were presented in the context of a linear scan over SUGRA-19 parameters, but remain
qualitatively valid in the case of a log scan over parameter space.
We also compare our results with those obtained in the minimal SUGRA or CMSSM model, where
wino-like $\tz_1$s don't appear. In the case of mSUGRA, the constrained GUT scale parameter
choices actually make neutralino CDM somewhat more likely than in the SUGRA-19 case.

If the strong CP problem is solved by the PQ mechanism in the context of SUSY models, then
an attractive alternative for dark matter appears: mixed axion/axino CDM.
A scan over SUGRA-19 parameters, augmented by four extra PQMSSM parameters, shows that the
mixed axion/axino DM abundance can be generated. While the measured CDM abundance doesn't lie at the
peak of the predicted $a\ta$ abundance, it also doesn't lie at a minimum. Therefore, in our viewpoint,
the mixed axion/axino particles seem a more plausible candidate for CDM.

In the end, the issue will have to be resolved by experiment. 
A thorough search for relic axions with $f_a/N\sim 10^9-10^{16}$ GeV is needed. Currently, only
a tiny portion of QCD axion parameter space has been explored\cite{admx}.
A major consequence of our analysis
then is to motivate our experimental colleagues to consider the possibility of new, more encompassing
probes of axion dark matter. These should probe much more broadly in axion mass, and
much more deeply in axion coupling.

%%%%%%%%%%%%%%%%%%%%%%%%%%%%%%%%%%%%%%%%%%%%%%%%
\acknowledgments
%%%%%%%%%%%%%%%%%%%%%%%%%%%%%%%%%%%%%%%%%%%%%%%%

We thank Xerxes Tata for reading this manuscript.
%This research was supported in part by the U.S. Department of Energy
%grant numbers DE-FG02-97ER41022.  
	
% ---- Bibliography ----
%


\begin{thebibliography}{99}
%
\bibitem{wmap7} E. Komatsu {\it et al.} (WMAP collaboration), 
arXiv:1001.4538 (2010).
%
\bibitem{dmreview} For reviews, see {\it e.g.}
G.~Jungman, M.~Kamionkowski and K.~Griest,\prep{267}{1996}{195};
A.~Lahanas, N.~Mavromatos and D.~Nanopoulos, \ijmpd{12}{2003}{1529};
M.~Drees, \hepph{0410113};
K.~Olive, ``Tasi Lectures on Astroparticle Physics'', \astroph{0503065};
G. Bertone, D. Hooper and J. Silk, \prep{405}{2005}{279}.
%
\bibitem{gauge} U. Amaldi, W. de Boer and H. Furstenau, \plb{260}{1991}{447};
J. Ellis, S. Kelley and D. V. Nanopoulos, \plb{260}{1991}{131};
P. Langacker and Luo, \prd{44}{1991}{817}.
%
\bibitem{drw} S. Dimopoulos, S. Raby and F. Wilczek, \prd{24}{1981}{1681};
S. Dimopoulos and H. Georgi, \npb{193}{1981}{150};
L. Ibanez and G. Ross, \plb{105}{1981}{439};
N. Sakai, \zpc{11}{1981}{153};
M. Einhorn and D. R. T. Jones, \npb{196}{1982}{475};
W. Marciano and G. Senjanovic, \prd{25}{1982}{3092}.
%
\bibitem{wss} H.~Baer and X.~Tata, {\it Weak Scale Supersymmetry: From 
Superfields to Scattering Events}, (Cambridge University Press, 2006).
%
\bibitem{pq} R. Peccei and H. Quinn, \prl{38}{1977}{1440} and
\prd{16}{1977}{1791}.
%
\bibitem{ww} S. Weinberg, \prl{40}{1978}{223};
F. Wilczek, \prl{40}{1978}{279}.
%
\bibitem{steffen_review} For recent reviews of axion/axino dark matter, see
F. Steffen, arXiv:0811.3347 (2008);
L. Covi and J. E. Kim, arXiv:0902.0769 (2009). 
%
\bibitem{msugra} A.~Chamseddine, R.~Arnowitt and P.~Nath, \prl{49}{1982}{970};
R.~Barbieri, S.~Ferrara and C.~Savoy, \plb {119}{1982}{343};
N.~Ohta, Prog. Theor. Phys. {\bf 70} (1983) 542; L. Hall,
J. Lykken and S. Weinberg, \prd {27}{1983}{2359}.
%
\bibitem{inoDM} H.~Goldberg, \prl {50}{1983}{1419};
J.~Ellis, J.~Hagelin, D.~Nanopoulos and M.~Srednicki, \plb{127}{1983}{233};
J.~Ellis, J.~Hagelin, D.~Nanopoulos, K.~Olive and M.~Srednicki, \npb{238}{1984}{453}.
%
\bibitem{bulk} P.~Nath and R.~Arnowitt, \prl {70}{1993}{3696}; 
A. Bottino, V. Alfaro, N. Fornengo, G. Mignola and M. Pignone, 
\app{2}{1994}{67};
H.~Baer and M.~Brhlik, \prd{53}{1996}{597};
V. Berezinsky, A. Bottino, J. Ellis, N. Fornengo, G. Mignola and S. Scopel, 
\app{5}{1996}{1};
V.~Barger and C.~Kao, \prd{57}{1998}{3131}.
%
\bibitem{stau} J.~Ellis, T.~Falk and K.~Olive, \plb{444}{1998}{367}; 
J.~Ellis, T.~Falk, K.~Olive and M.~Srednicki, \app{13}{2000}{181};
M.E.~G\'{o}mez, G.~Lazarides and C.~Pallis, \prd{61}{2000}{123512}
and \plb{487}{2000}{313};
A.~Lahanas, D.~V.~Nanopoulos and V.~Spanos, \prd{62}{2000}{023515};
R.~Arnowitt, B.~Dutta and Y.~Santoso, \npb{606}{2001}{59}; 
see also Ref.~\cite{isared}.
%
\bibitem{hb_fp} K.~L.~Chan, U.~Chattopadhyay and P.~Nath, \prd{58}{1998}{096004};
J.~Feng, K.~Matchev and T.~Moroi, \prl{84}{2000}{2322} and 
\prd{61}{2000}{075005}; see also 
H.~Baer, C.~H.~Chen, F.~Paige and X.~Tata, \prd{52}{1995}{2746} and 
\prd{53}{1996}{6241}; 
H.~Baer, C.~H.~Chen, M.~Drees, F.~Paige and X.~Tata, \prd{59}{1999}{055014}; 
for a model-independent approach, see
H.~Baer, T.~Krupovnickas, S.~Profumo and P.~Ullio, \jhep{0510}{2005}{020}.
%
\bibitem{Afunnel} M.~Drees and M.~Nojiri, \prd{47}{1993}{376}; 
H.~Baer and M.~Brhlik, \prd{57}{1998}{567};
H.~Baer, M.~Brhlik, M.~Diaz, J.~Ferrandis, P.~Mercadante,
P.~Quintana and X.~Tata, \prd{63}{2001}{015007};
J.~Ellis, T.~Falk, G.~Ganis, K.~Olive and M.~Srednicki, \plb{510}{2001}{236}; 
L.~Roszkowski, R.~Ruiz de Austri and T.~Nihei, \jhep{0108}{2001}{024}; 
A.~Djouadi, M.~Drees and J.~L.~Kneur, \jhep{0108}{2001}{055}; 
A.~Lahanas and V.~Spanos, \epjc{23}{2002}{185}.
%
\bibitem{higgs} R.~Arnowitt and P.~Nath, \prl{70}{1993}{3696};
H.~Baer and M.~Brhlik, Ref.~\cite{inoDM};  
A.~Djouadi, M.~Drees and J.~Kneur, \plb{624}{2005}{60}.
%
\bibitem{eo} J. Ellis and K. Olive, \plb{514}{2001}{114}.
%
\bibitem{bbox} H. Baer and A. Box, arXiv:0910.0333 (2009).
%
\bibitem{bbs} H. Baer, A. Box and H. Summy, \jhep{0908}{2009}{080}.
%
\bibitem{axmass} E. J. Chun, J. E. Kim and H. P. Nilles, 
\plb{287}{1992}{123}; E. J. Chun and A. Lucas, \plb{357}{1995}{43}. 
%
\bibitem{bghr} C. Berger, J. Gaines, J. Hewett and T. Rizzo, \jhep{0902}{2009}{023}.
%
\bibitem{py} S. Profumo and C. E. Yaguna, \prd{70}{2004}{095004}.
%
\bibitem{aaqfh} S. S. AbdusSalam, B. C. Allanach, F. Quevedo, F. Feroz and M. Hobson, arXiv:0904.2548 (2009).
%
\bibitem{bbps} G. Belanger, F. Boudjema, A. Pukhov and R. K. Singh, 
\jhep{0911}{2009}{026}.
%
\bibitem{isajet} F. Paige, S. Protopopescu, H. Baer and X. Tata, \hepph{0312045}; 
http://www.nhn.ou.edu/$\sim$isajet/
%
\bibitem{bfkp} H.~Baer, J.~Ferrandis, S.~Kraml and W.~Porod, \prd{73}{2006}{015010}.
%
\bibitem{mv} S. Martin and M. Vaughn, \prd{50}{1994}{2282}.
%
\bibitem{haber} H. E. Haber, R. Hempfling and A. Hoang, \zpc{75}{1996}{539}.
%
\bibitem{kraml} G. Belanger, S. Kraml and A. Pukhov, \prd{72}{2005}{015003}.
%
\bibitem{isared} H. Baer, C. Balazs, A. Belyaev, \jhep{0203}{2002}{042}.
%
\bibitem{ggsy} G. Gelmini, P. Gondolo, A. Soldatenko and C. Yaguna, 
\prd{74}{2006}{083514}.
%
%\bibitem{bbefms} 
%
\bibitem{wtn} N. Arkani-Hamed, A. Delgado and G. F. Giudice, \npb{741}{2006}{108};
H.~Baer, A.~Mustafayev, E.~Park and X.~Tata,
JCAP {\bf 0701}, 017 (2007); H.~Baer, A.~Mustafayev, E.~Park and X.~Tata, \jhep
  {0805}{2008}{058}.
%
\bibitem{ewft} 
R. Barbieri and G. Giudice, \npb{306}{1988}{63};
G. Anderson and D. Castano, \plb{347}{1995}{300} and \prd{52}{1995}{1693}. 
%
\bibitem{ksvz} J. E. Kim, \prl{43}{1979}{103};
M. A. Shifman, A. Vainstein and V. I. Zakharov, \npb{166}{1980}{493}.
%
\bibitem{dfsz} M. Dine, W. Fischler and M. Srednicki, \plb{104}{1981}{199};
A. P. Zhitnitskii, \sjp{31}{1980}{260}.
%
\bibitem{gravproblem} S. Weinberg, \prl{48}{1982}{1303};
R. H. Cyburt, J. Ellis, B. D. Fields and K. A. Olive, \prd{67}{2003}{103521};
K. Jedamzik, \prd{70}{2004}{063524};
M. Kawasaki, K. Kohri and T. Moroi, \plb{625}{2005}{7}.
%
\bibitem{rtw} K. Rajagopal, M. Turner and F. Wilczek, 
\npb{358}{1991}{447}.
%
\bibitem{jlm} K. Jedamzik, M. LeMoine and G. Moultaka,
JCAP{\bf 0607} (2006) 010.
%
\bibitem{bbn} K. Kohri, T. Moroi and A. Yotsuyanagi, \prd{73}{2006}{123511};
for an update, see
M. Kawasaki, K. Kohri, T. Moroi and A. Yotsuyanagi, \prd{78}{2008}{065011};
see also J.~Pradler and F.~D.~Steffen, \plb{648}{2007}{224}.
%
\bibitem{kcreview} For recent reviews, 
see P. Sikivie, \hepph{0509198}; M. Turner, \prep{197}{1990}{67}; 
J. E. Kim and G. Carosi, arXiv:0807.3125 (2008);
%
\bibitem{ckkr} L. Covi, J. E. Kim and L. Roszkowski, \prl{82}{1999}{4180}; 
L. Covi, H. B. Kim, J. E. Kim and L. Roszkowski, \jhep{0105}{2001}{033}.
L. Covi, L. Roszkowski and Small, \jhep{0207}{2002}{023}. 
%
\bibitem{steffen} A. Brandenburg and F.~Steffen,
JCAP{\bf 0408} (2004) 008.
%
\bibitem{strumia} A. Strumia, arXiv:1003.5847 (2010).
%
\bibitem{absik} L. F. Abbott and P. Sikivie, \plb{120}{1983}{133};
J. Preskill, M. Wise and F. Wilczek, \plb{120}{1983}{127};
M. Dine and W. Fischler, \plb{120}{1983}{137};
M. Turner, \prd{33}{1986}{889}.
%
\bibitem{vg} L. Visinelli and P. Gondolo, 
\prd{0903}{2009}{035024}.
%
\bibitem{ntlepto} G. Lazarides and Q. Shafi, \plb{258}{1991}{305};
K. Kumekawa, T. Moroi and T. Yanagida, \ptp{92}{1994}{437};
T. Asaka, K. Hamaguchi, M. Kawasaki and T. Yanagida, \plb{464}{1999}{12}.
%
\bibitem{admx} 
L. Duffy {\it et al.}, \prl{95}{2005}{091304} and \prd{74}{2006}{012006};
for a review, see S. Asztalos, L. Rosenberg, K. van Bibber, P. Sikivie
and K. Zioutas, \arnps{56}{2006}{293}.
%
\end{thebibliography}
\end{document}